\documentclass[aps,prb,twocolumn,superscriptaddress,showpacs,amsmath,amssymb]{revtex4}

\usepackage{graphicx}%  Include figure files
\usepackage[floatfix]{epsfig}%  Include figure files
\usepackage{bm}% bold math

\begin{document}

%\preprint{}

\title{Change of quasiparticle dispersion in crossing $T_c$ in the underdoped cuprates}


\author{T.~Eckl}
\email{eckl@physik.uni-wuerzburg.de}
\affiliation{Institut f\"ur Theoretische Physik und Astrophysik,
Universit\"at W\"urzburg, Am Hubland, D-97074 W\"urzburg, Germany}

\author{W.~Hanke}
%\email{hanke@physik.uni-wuerzburg.de}
\affiliation{Institut f\"ur Theoretische Physik und Astrophysik,
Universit\"at W\"urzburg, Am Hubland, D-97074 W\"urzburg, Germany}

\author{S. V.~Borisenko}
\email{S.Borisenko@ifw-dresden.de}
\affiliation{Institute for Solid State Research, IFW-Dresden,
P.~O.~Box 27 00 16, D-01171 Dresden, Germany}

\author{A. A.~Kordyuk}
\affiliation{Institute for Solid State Research, IFW-Dresden,
P.~O.~Box 27 00 16, D-01171 Dresden, Germany}
\affiliation{Institute of Metal Physics of National Academy of Sciences of Ukraine, 03142 Kyiv, Ukraine}

\author{T.~Kim}
\affiliation{Institute for Solid State Research, IFW-Dresden,
P.~O.~Box 27 00 16, D-01171 Dresden, Germany}

\author{A.~Koitzsch}
\affiliation{Institute for Solid State Research, IFW-Dresden,
P.~O.~Box 27 00 16, D-01171 Dresden, Germany}

\author{M.~Knupfer}
\affiliation{Institute for Solid State Research, IFW-Dresden,
P.~O.~Box 27 00 16, D-01171 Dresden, Germany}

\author{J.~Fink}
%\email{J.Fink@ifw-dresden.de}
\affiliation{Institute for Solid State Research, IFW-Dresden,
P.~O.~Box 27 00 16, D-01171 Dresden, Germany}

\date{\today}

\begin{abstract}

One of the most remarkable properties of the high-temperature
superconductors is a pseudogap regime appearing in the underdoped 
cuprates above the superconducting transition temperature $T_c$.
The pseudogap continously develops out of the superconducting gap.
In this paper, we demonstrate by means of a detailed comparison
between theory and experiment that the characteristic
change of quasiparticle dispersion in crossing $T_c$ in the underdoped cuprates
can be understood as being due to phase fluctuations of the
superconducting order parameter. 
In particular, we show that within a phase fluctuation model 
the characteristic back-turning BCS bands disappear above $T_c$ whereas the gap remains open.
Furthermore, the pseudogap rather has a U-shape instead of the 
characteristic V-shape of a $d_{x^2-y^2}$-wave pairing symmetry and starts closing from
the nodal $\vec{k}=(\frac{\pi}{2},\frac{\pi}{2})$ directions, whereas it
rather fills in at the anti-nodal
$\vec{k}=(\pi,0)$ regions, yielding further support to the  phase
fluctuation scenario.

\end{abstract}

\pacs{71.10.Fd, 71.27.+a, 74.25.Jb, 74.72.-h, 79.60.-i}

\maketitle

\section{Introduction}

More than 15 years after the discovery of the high-$T_c$
superconductors, the mechanisms leading to their unusual 
properties are still under debate. Especially the
pseudogap phase, which appears in various experiments below 
a characteristic temperature $T^*$ in the underdoped region 
of the phase diagram as a reduction of spectral weight 
\cite{ku.fi.01,ec.sc.02}, might be a key to a better 
understanding of the  high-$T_c$ superconducting (SC) cuprates.
In 1995 Emery and Kivelson \cite{em.ki.95} proposed, that the proximity
to the Mott insulating phase implies a strongly reduced phase
stiffness $J \sim \frac{\rho_s(0)}{m^*}$ compared to the usual BCS case.
This causes the phase ordering temperature $T_{\varphi} \sim J$ to be much lower
than the mean-field pair-binding temperature $T_c^{MF}$.
Taking this idea one step further \cite{em.ki.95}
implies that at least part of the pseudogap behavior might 
be due to a kind of ``pre-formed'' Cooper pairs
which form at a temperature $T^* \equiv T_c^{MF}$ well above the
actual SC transition temperature $T_c \equiv T_{\varphi}$, where
phase coherence among these pairs finally sets in.
This phase fluctuation scenario also explains quite natural
the strongly enhanced Nernst signal above $T_c$ 
in the underdoped cuprates \cite{wa.xu.01}.

In previous work, we have already shown that indeed a two-dimensional 
BCS-like Hamiltonian with a $d_{x^2-y^2}$-wave gap and phase
fluctuations, which were treated by a Monte-Carlo simulation of an
$XY$ model, yields results
which compare very well with scanning tunneling measurements 
over a wide temperature range \cite{ec.sc.02,ku.fi.01}. Furthermore,
this phenomenological phase fluctuation model
was also able to explain the possible ``violation'' of the
in-plane optical integral in underdoped Bi$_2$Sr$_2$CaCu$_2$O$_{8+\delta}$ (Bi2212) \cite{ec.ha.03}.

However, for the phase fluctuation description to be correct 
over a wide temperature range, one needs
a mechanism that produces ``cheap'' vortices,
so that the only energy scale is the stiffness $J$ 
and the dominating fluctuation channel is that 
of the phase of the SC order parameter. Mechanisms that can lead to 
a small vortex core energy range from the more conventional 
picture of a granular superconductor, where the vortices arrange 
themselves to live in the insulating regions between the SC grains, 
up to the existence of a competing order that exists inside 
the vortex cores. As soon as the superconducting order parameter
is suppressed inside the vortex core, the system develops the
competing order instead of going into a normal conducting paramagnetic
state and thus has a much smaller vortex energy compared to
a conventional BCS superconductor. Recently, it was shown \cite{ho.le.03u},
that vortices with staggered-flux core can provide a way to understand
the low vortex energy over a wide temperature range above $T_c$.
In all cases, the small phase stiffness and the low vortex core energy
have the same origin, which is the proximity to the Mott-insulating state.

In this paper, we present theoretical results 
on the quasiparticle dispersion, which --- when compared with
experimental data --- give a clear fingerprint towards
a possible phase fluctuation scenario for the origin of the pseudogap. Earlier angle 
resolved photoemission (ARPES) results have shown deviations from the 
simple BCS $d_{x^2-y^2}$-wave form of the SC gap in underdoped Bi2212 
\cite{me.no.99,bo.ko.02}, which might be due to a change in the pairing
interaction in the proximity of the AF insulating phase. By analyzing
the temperature dependence of the quasiparticle (QP) dispersion, we want
to show, that the change of the QP dispersion in crossing
$T_c$ from the SC to the pseudogap region can be understood quite
naturally by the assumption that the pseudogap is caused by phase 
fluctuations of the SC gap. 
Moreover, the phase fluctuation scenario
also explains the deviations from the simple $d_{x^2-y^2}$-wave form of 
SC and pseudogap \cite{no.di.98} in the underdoped cuprates.
Using the ARPES with tunable excitation photon energy we disentangle bilayer splitting related effects and determine the true dispersion and the leading edge gap (LEG) function corresponding to the bonding band in the pseudogap regime of Pb-Bi2212.

Since below $T_c$ the QP excitations are perfectly BCS-like unless in the
extremely underdoped region \cite{ma.sa.03}, it is tempting to start from the 
BCS ground state and see how it is destroyed by including phase
fluctuations \cite{fr.mi.98,kw.do.99,fr.te.01,herb.02}.

\section{Model and calculations}

In the following we use a phenomenological phase fluctuation model
which has already been shown to successfully account for
the pseudogap observed in tunneling experiments \cite{ec.sc.02} and
which was also able to explain the possible ``violation'' of the
in-plane optical integral in underdoped Bi2212 \cite{ec.ha.03}. We consider the Hamiltonian
\begin{equation}
H=H_0+H_1,
\label{one}
\end{equation}
where $H_0$ is the usual tight-binding Hamiltonian of
non-interacting electrons on a two-dimensional (2D) square lattice
\begin{equation}
H_0=-t \sum_{\langle
  \vec{i}\,\vec{j}\rangle,\sigma}(c^\dagger_{\vec{i}\,\sigma}c_{\vec{j}\,\sigma}+c^\dagger_{\vec{j}\,\sigma}c_{\vec{i}\,\sigma})
- \mu \sum_{\vec{i},\sigma} n_{\vec{i}\,\sigma}.
\label{ham_0}
\end{equation}
Here, $c^\dagger_{\vec{i}\,\sigma}$ ($c_{\vec{i}\,\sigma}$) creates
(annihilates) an electron of spin $\sigma$ on the $\vec{i}^{\, th}$
site of the 2D square lattice and  $n_{\vec{i}\,\sigma}=c^\dagger_{\vec{i}\,\sigma}c_{\vec{i}\,\sigma}$
is the number operator. $t$  denotes an effective
nearest-neighbor hopping-term
and $\mu$ is the chemical potential.
The angles $\langle \cdots \rangle$ 
indicate sums over nearest-neighbor sites of the 2D square lattice.

The second part of the Hamiltonian $H_1$ contains a BCS-like $d$-wave interaction,
which is given by
\begin{equation}
H_1=-g\sum_{\vec{i}\,\vec{\delta}}(\Delta_{\vec{i}\,\vec{\delta}}\langle\Delta_{\vec{i}\,\vec{\delta}}^\dagger\rangle 
+\Delta_{\vec{i}\,\vec{\delta}}^\dagger\langle\Delta_{\vec{i}\,\vec{\delta}}\rangle),
\label{ham_i_d}
\end{equation}
with $\vec{\delta}$ connecting nearest-neighbor sites. 
The coupling constant $g$ stands for the strength of the effective
next-neighbor $d_{x^2-y^2}$-wave pairing-interaction.
The origin of this pairing interaction is unimportant for the
further calculation. It can be either of pure electronic origin,
like spin fluctuations, or phonon mediated. The only
important thing is, that there exists an effective pairing
interaction, that produces a finite local $d_{x^2-y^2}$-wave gap as
one goes below a certain temperature $T^*$.
In contrast to conventional BCS theory, we consider the pairing-field amplitude
not as a constant real number, but rather as a complex number
\begin{equation}
\langle\Delta_{\vec{i}\,\vec{\delta}}^\dagger\rangle=
\frac{1}{\sqrt{2}}\langle c_{\vec{i}\,\uparrow}^\dagger c_{\vec{i}+\vec{\delta}\,\downarrow}^\dagger
-c_{\vec{i}\,\downarrow}^\dagger c_{\vec{i}+\vec{\delta}\,\uparrow}^\dagger\rangle=
\Delta\,e^{i \,\Phi_{\vec{i} \vec{\delta}}},
\label{pair1}
\end{equation}
with a {\sl constant} magnitude $\Delta$ and a {\sl fluctuating} bond-phase field $\Phi_{\vec{i} \vec{\delta}}$.
In order to get a description, where the {\sl center of mass}
phases of the Cooper pairs are the only relevant degrees of freedom \cite{pa.ra.00},
we write the $d_{x^2-y^2}$-wave bond-phase field in the following way
\begin{equation}
\Phi_{\vec{i} \vec{\delta}}=\left\{\begin{array}{l@{\quad \mathrm{for} \quad}l}
(\varphi_{\vec{i}} + \varphi_{\vec{i}+\vec{\delta}})/2 & \text{$\vec{\delta}$ in $x$-direction} \\
(\varphi_{\vec{i}} + \varphi_{\vec{i}+\vec{\delta}})/2 +\pi & \text{$\vec{\delta}$ in $y$-direction,} 
\end{array} \right.
\label{pair2}
\end{equation}
where $\varphi_{\vec{i}}$ is the {\sl center of mass} phase of a Cooper
pair localized at lattice site $\vec{i}$. 

In order to account for the proximity to the Mott insulating state
and thus the low superfluid density, we perform a {\sl quenched average}
over all possible phase configurations with the statistical weight
given by the classical $XY$ free energy
\begin{equation}
F\left[\varphi_i\right] = - J \sum_{\langle ij\rangle}
\cos\left(\varphi_i-\varphi_j\right),
\label{two}
\end{equation}
where the phase stiffness $J$ determines the
Berezinskii-Kosterlitz-Thouless
transition temperature $T_{BKT}$ to a quasi phase ordered
state which we take as $T_c$. The $XY$ free energy is defined on a 
coarse-grained lattice with the \emph{scale} of the 
lattice spacing given by the pair coherence length \cite{pa.ra.00} $\xi_0 \sim  \frac{v_F}{\pi \Delta}$
. Now, the underdoped cuprates
are in an intermediate coupling regime between large
BCS mean-field pairs and tightly-bound BEC pairs \cite{pa.ca.99,mi.ro.99}, 
with the pair-size coherence length $\xi_0$ given by
3 to 4 times the basic Cu-Cu lattice spacing.
For a typical $36 \times 36 $ fermionic lattice, which is numerical
feasible, we would only have a $9 \times 9$ phase lattice on top of it.
This would not allow for any proper temperature scaling of the phase
correlation length $\xi(T)$ and obscure the Kosterlitz-Thouless transition.
Therefore we have chosen to set $\Delta_{sc}=1.0 \,t$.
This yields $\xi_0 \lesssim 1$ and allows the Monte Carlo phase
simulation to be carried out on the same $L\times L$ lattice that is
used for the diagonalization of the fermionic Hamiltonian \cite{ec.ha.03}. 
In addition, the choice of $\Delta_{sc}=1.0\,t$ automatically
introduces the important short distance cut-off.
Finally we set  $ T_c \approx \frac{1}{4} T^*$, where we had the STM experiments
of Ref.~\onlinecite{ku.fi.01} in mind.

\section{Experimental details}

The ARPES experiments were carried out using angle-multiplexing electron energy analyzers. Spectra were recorded either with $h\nu=$21.218 eV photons from a He source or using radiation from the U125/1-PGM beamline at the BESSY synchrotron radiation facility. The total energy resolution was set to 17 meV (FWHM) at $h\nu=$ 38 eV. The angular resolution was kept below 0.2$^\circ$ both along and perpendicular to the analyzer entrance slit. Data shown in Fig.~\ref{fig4} were taken with 0.2$^\circ$ x 0.3$^\circ$ angular resolution. The data were collected on two similar underdoped, modulation-free single crystals of Pb-Bi2212 ($T_c$=77K).

\section{Discussion of results}

\subsection{Dispersion}

\begin{figure}[t]
\begin{center}
\epsfig{file=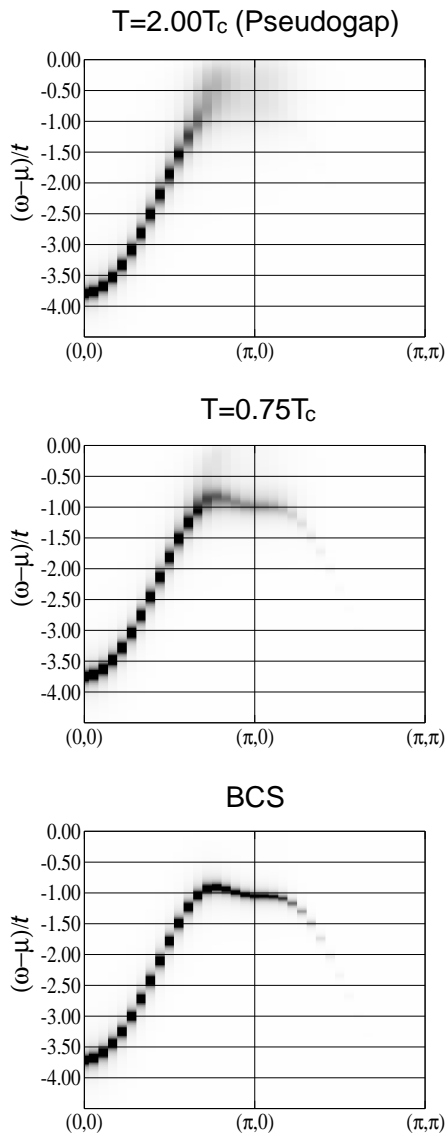,width=6cm}
\end{center}
\caption[]{Spectral weight $A(\vec{k},\omega)$ in the pseudogap state
($T=2.0\,T_c$, top) and in the superconducting state slightly below $T_c$
($T=0.75\,T_c$) calculated from the phase fluctuation model.
For comparison we also show the spectral weight for the phase coherent
BCS limit (bottom).}
\label{fig1}
\end{figure}

\begin{figure}[t]
\begin{center}
\epsfig{file=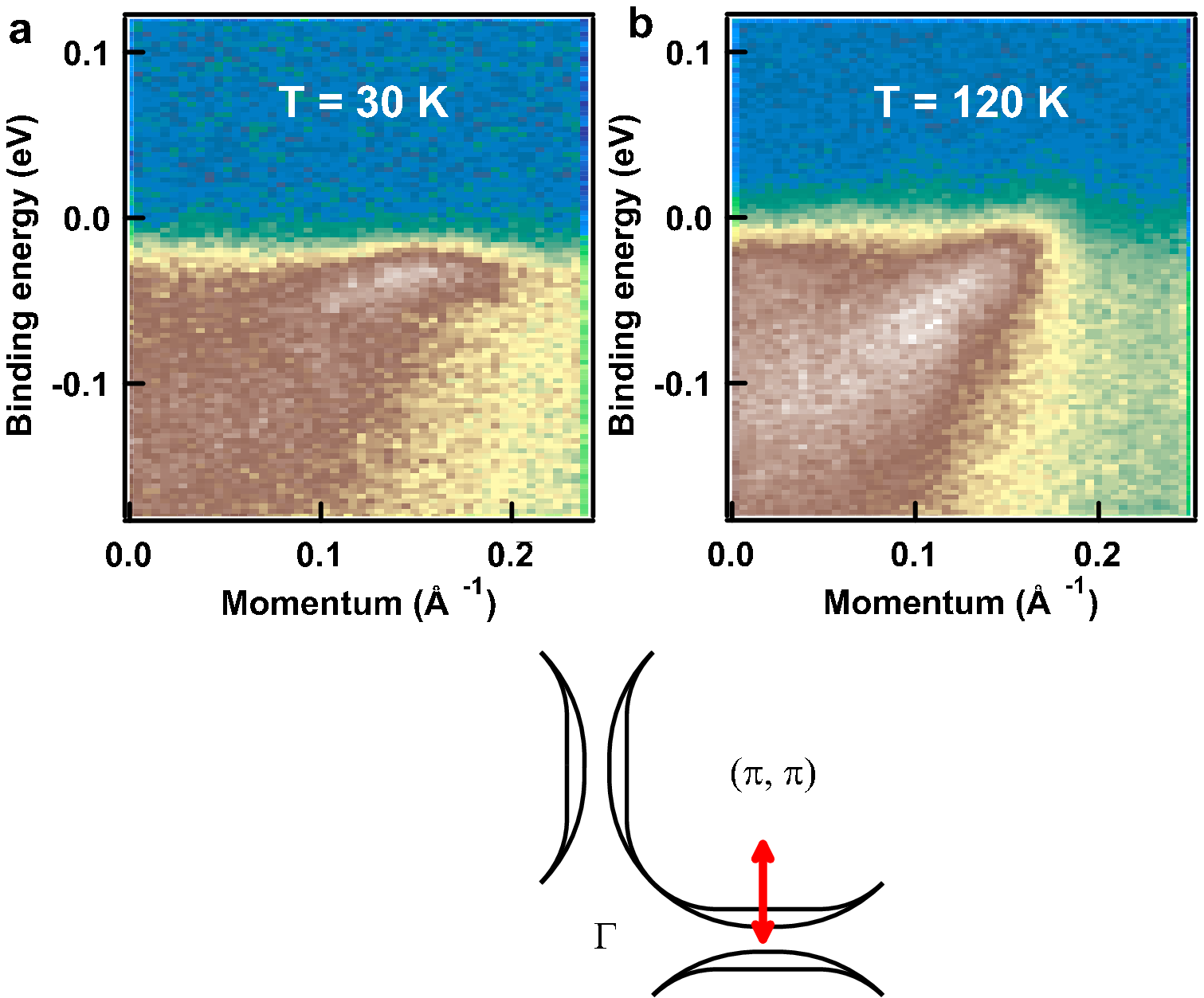,width=8.8cm}
\end{center}
\caption[]{a) Superconducting state. Energy distribution of the
  photoemission intensity along the direction shown as a red arrow on
  the sketch below. The BCS-like dispersion is clearly observed for
  the bonding band. b) Pseudogap state. No more bending back of the
  dispersion is observed. Instead, spectral weight fades upon approaching the Fermi level.}
\label{fig2}
\end{figure}

Fig.~\ref{fig1} shows the quasiparticle dispersion calculated from our 
phenomenological phase fluctuation model for $10\%$ doping
($\langle n \rangle = 0.9$). The spectral weight is plotted
along the $(0,0)\rightarrow(\pi,0)\rightarrow(\pi,\pi)$
direction through the Brillouin zone (BZ). The free dispersion
would cross the Fermi surface close to the $(\pi,0)$-point. 
One can clearly see that the characteristic (back-turning)
Bogoliubov quasiparticle band disappears in the pseudogap state 
above $T_c$. Instead, one obtains a sharp 
quasiparticle dispersion which runs straight towards the 
Fermi energy and than fades out at a distance of the order of
the superconducting (SC) gap $\Delta_{sc}$.
This is in complete agreement with the experimentally 
observed dispersion in underdoped Bi2212 which is shown in Fig.~\ref{fig2}. 

The angle resolved data presented in Fig.~\ref{fig2} provide an insight into how the pseudogap is actually created near $k_f$. In the superconducting state a characteristic BSC-like back-dispersion is easily seen. This clarity is achieved by the careful choice of the excitation photon energy. Exactly near h$\nu$=38 eV the emission probability for the bonding band is much higher than for the antibonding band \cite{ko.bo.02o,bo.ko.03a} and bilayer-related complications are thus avoided. Above $T_c$ in the pseudogap state the characteristic BCS behavior is replaced by the straight dispersion and strong depletion of the spectral weight towards $E_f$, which, as will be shown below, still leaves the energy gap in the spectrum.

Furthermore, Fig.~\ref{fig1} shows that the sharp quasiparticle
features close to the $(\pi,0)$-point are getting lost above
$T_c$ within the phase fluctuation model. The sharp coherent $(\pi,0)$-peaks dissappear
and broad incoherent weight {\sl fills-in} the gap.
Exactly this behavior was observed before in photoemission studies of the
pseudogap \cite{no.di.98,fe.lu.00,no.ka.01,fe.da.02,sa.ma.02}
and is also responsible for the characteristic temperature  
dependence of the scanning tunneling gap in the underdoped cuprates
\cite{ec.sc.02,ku.fi.01}, where the pseudogap fills-in instead of closing.
Interestingly, not only SC fluctuations \cite{pi.pi.03u}, but also staggered
flux fluctuations \cite{ho.le.03} can lead to this temperature dependence of
the $(\pi,0)$-photoemission-peak.

The disappearance of the characteristic BCS bands above $T_c$ within
the phase fluctuation picture can be understood 
by the fact, that the BCS wave-function is a coherent superposition
of wave functions with different number of electron pairs \cite{tinkhamN}
\begin{equation}
|\Psi_{BCS}\rangle = \prod_k(u_k+v_k \, c_{k\,\uparrow}^\dagger c_{-k\,\downarrow}^\dagger)|\phi_0\rangle
=\sum_N \lambda_N |\Psi_N\rangle,
\end{equation}
where $|\Psi_N\rangle$ is a N-particle wavefunction.
The quantum mechanical uncertainty in the particle number is given by:
\begin{equation}
(\Delta N)^2 =4 \sum_k u^2_k v^2_k.
\label{number}
\end{equation}
Now $v_k^2=1-u_k^2$ is the momentum distribution function for $T=0$ and the weight of a quasiparticle peak
at momentum $k$ is given by $v_k^2$ ($u_k^2$) for $E<E_f$ ($E>E_f$). 

In the normal metallic state with $\Delta=0$, one gets a sharp cut-off in  $v_k^2$ ($u_k^2$) at 
the Fermi-wavevector $k=k_f$ so that $(\Delta N)^2  \equiv 0$. In the BCS-superconducting
state, however, $v_k^2$ ($u_k^2$) are finite also beyond the Fermi-wavevector $k_f$,
which means that also for $k>k_f$  ($k<k_f$) one gets spectral weight at  $E<E_f$ ($E>E_f$).
This produces the characteristic BCS band structure, with bands approaching $E_f$
from below (above) and then turning back to higher binding (quasiparticle) energies. 

Now what happens if one introduces an arbitrary phase factor into the
BCS wave function \cite{tinkhamN}
\begin{equation}
|\Psi_{\varphi}\rangle = \prod_k(|u_k|+|v_k| e^{i\varphi} \, c_{k\,\uparrow}^\dagger c_{-k\,\downarrow}^\dagger)|\phi_0\rangle.
\end{equation}
Integrating over all possible phases yields \cite{tinkhamN}
\begin{equation}
\begin{split}
|\Psi_{N}\rangle &= \int^{2\pi}_0 d\varphi \, e^{-i N \varphi /2 }\prod_k(|u_k|+|v_k| e^{i\varphi} \, c_{k\,\uparrow}^\dagger 
c_{-k\,\downarrow}^\dagger)|\phi_0\rangle\\
&=\int^{2\pi}_0  d\varphi \, e^{-i N \varphi /2 } |\Psi_{\varphi}\rangle.
\end{split}
\end{equation}
This means, that one projects into an exact particle-number eigenstate
by making the relative phase of the Cooper-pairs completely uncertain.
Eq.~(\ref{number}) is a special case of the general uncertainty relation between
phase and particle number
\begin{equation}
\Delta N \, \Delta \varphi \gtrsim 1.
\end{equation}
The above described behavior corresponds to what is happening 
in the phase-fluctuation model as a function of temperature. 
Starting from a phase coherent state at $T=0$ with $\Delta \varphi
=0$, the particle number is completely uncertain with  $\Delta N$
given by Eq.~\ref{number}.
With increasing temperature, one {\sl gradually projects into a state
  with exact particle number} $N$.
In the temperature range, where the phases are completely uncorrelated
($\xi \sim \xi_0$), one then obtains $\Delta N =0$, and the 
back-turning BCS-bands must completely dissappear 
(loose weight for $k > k_f$, as seen in Figs.~\ref{fig1} and \ref{fig2}).
At finite temperatures, this situation corresponds to
a classical grand canonical average over ensembles with different number of particles, where each state has a
well defined particle number and is no longer a coherent quantum-mechanical superposition
of states with different number of particles.
Thus, we obtain a crossover from a BCS-like phase-ordered band structure
to a completely new phase-disordered {\sl pseudo-gaped} band structure.

\subsection{Superconducting gap and pseudogap}

\begin{figure}[t]
\begin{center}
\epsfig{file=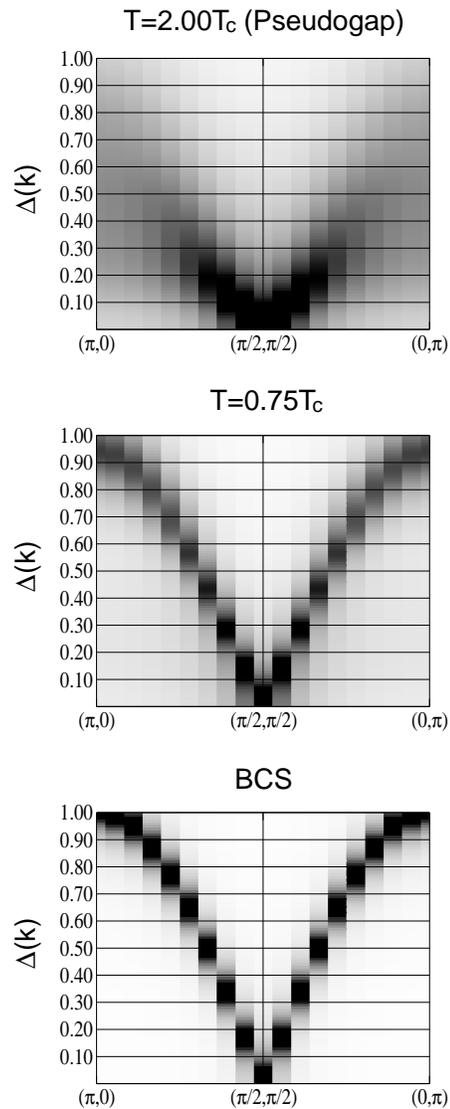,width=6cm}
\end{center}
\caption[]{Gap function $\Delta(\vec{k})$ in the pseudogap state
($T=2.0\,T_c$, top) and in the superconducting state slightly below $T_c$
($T=0.75\,T_c$) calculated from the phase fluctuation model.
For comparison we also show the BCS gap function (bottom).}
\label{fig3}
\end{figure}

\begin{figure}[t]
\begin{center}
\epsfig{file=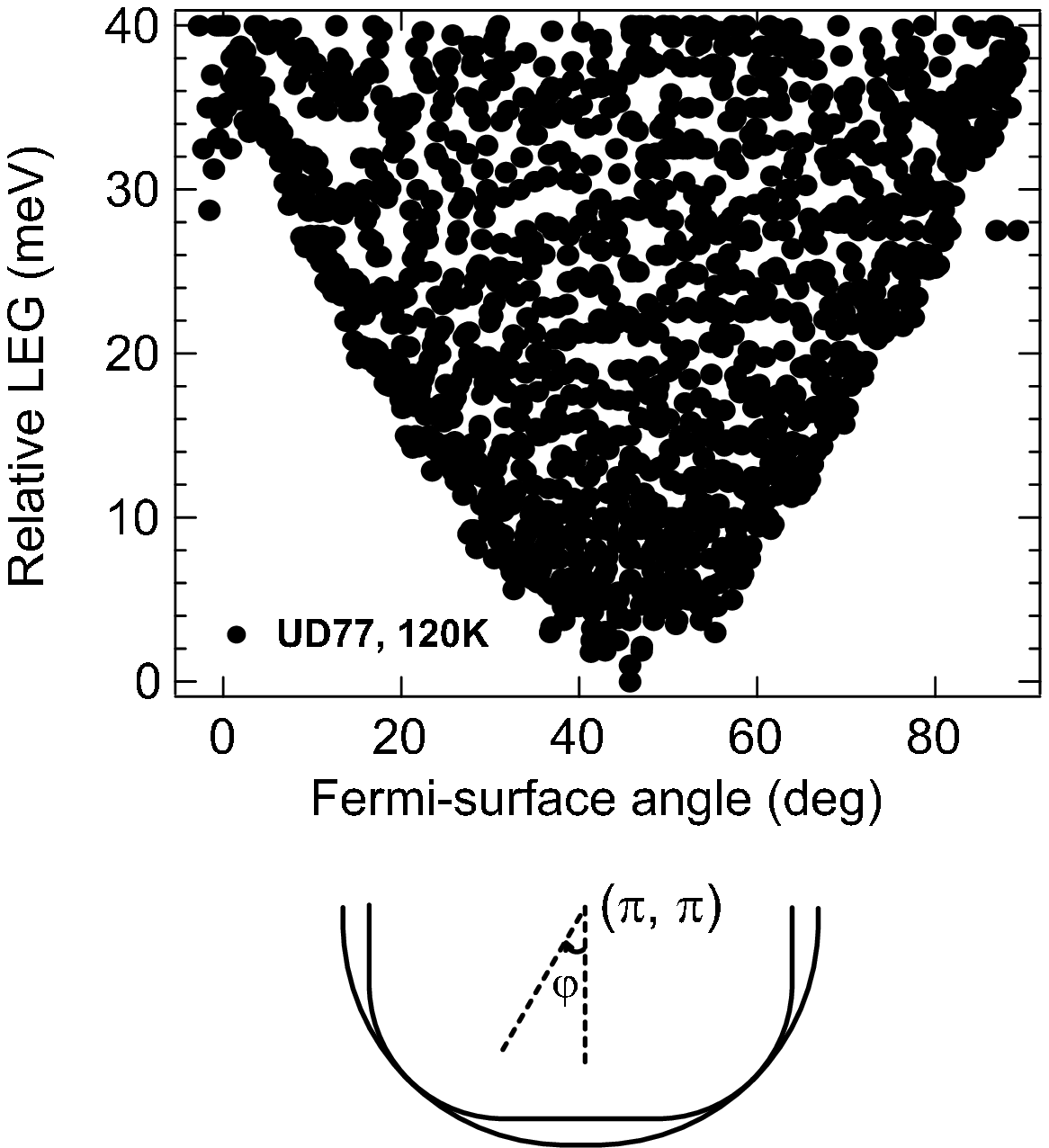,width=7.5cm}
\end{center}
\caption[]{Values of the leading edge (pseudo-)gap (LEG) as a function of the Fermi surface angle within the quadrant of the Brillouin zone (see lower panel) given with respect to the binding energy of the leading edge of a nodal energy distribution curve. The curve joining the low-gap extremity of these data points would represent the $\vec{k}$-dependence of the pseudogap. For details see Ref.~\onlinecite{bo.ko.02}.}
\label{fig4}
\end{figure}

Next we want to elucidate the effect of phase fluctuations on
the $\vec{k}$-dependence of the quasiparticle pairing gap. 
Therefore, we have plotted in Fig.~\ref{fig3} the 
quasiparticle dispersion obtained from MC simulations of the phase
fluctuation model along the Fermi surface of the free 
dispersion $\epsilon(\vec{k})$ at half-filling ($\langle n \rangle = 1.0$).
This gives us effectively the gap function $\Delta(\vec{k})$.
As can be seen in Fig.~\ref{fig3},
below $T_c$ one obtains the characteristic $V$-shape
of a gap with $d_{x^2-y^2}$ pairing symmetry.
As the temperature is raised, the quasiparticles
peaks are getting broader. 
In the pseudogap state above $T_c$, the spectral 
weight is getting rather incoherent close the
the $\vec{k}=(\pi,0)$ as was pointed out before.  
However, close to the nodal point of the gap function one still 
obtains a sharp quasiparticle dispersion. 
There, one can clearly see spectral weight 
shifting to lower binding energies
which produces an {\sl extended gapless region} 
in the pseudogap state close to
$\vec{k}=(\frac{\pi}{2},\frac{\pi}{2})$ 
instead of the nodal point in the superconducting
state below $T_c$. This behavior is in complete agreement with 
photoemission experiments \cite{no.di.98,sa.ma.02} 
which show that the pseudogap starts closing from 
$\vec{k}=(\frac{\pi}{2},\frac{\pi}{2})$ where 
one obtains a finite Fermi-arc  
but rather {\sl fills in} at $\vec{k}=(\pi,0)$
exactly as in Fig.~\ref{fig3} (top).

Furthermore, the pseudogap $\Delta(\vec{k})$ obtained from 
phase fluctuations of the local $d_{x^2-y^2}$ pairing-amplitude rather 
has a U-like shape (see Fig.~\ref{fig3}, top) than the 
characteristic V-shape of a BCS $d_{x^2-y^2}$-gap.
 For comparison, Fig.~\ref{fig4} shows the experimentally
observed pseudogap in underdoped Bi2212. 
One can clearly see that the experimentally determined
pseudogap has exactly the U-like form
that we have obtained from the phase fluctuation model.
This deviation from the pure $d_{x^2-y^2}$-wave form was 
also observed in the superconducting state 
of very underdoped cuprates
and interpreted as higher harmonic contributions to the pairing
function \cite{me.no.99,bo.ko.02}.
However, these experimental results just might indicate the possible relevance of 
quantum phase fluctuations in this region of the phase diagram. 
Our results on the effects of phase fluctuations on the form of the  
pairing gap could also be of some relevance for electron doped
cuprates \cite{al.me.99} where a possible crossover from a $d_{x^2-y^2}$ (or anisotropic $s$-wave) to
a pure $s$-wave symmetry of the superconducting gap as a function 
of electron doping was observed
\cite{bi.fo.02,sk.ki.02}.

\section{Summary and conclusion}

In conclusion, we have elaborated 
the important role that phase fluctuation effects 
might play in the underdoped cuprates.
With a detailed comparison between theory and experiment
we were able to show how phase fluctuations influence the quasiparticle spectra.
In particular the disappearance of the BCS-Bogoliubov
quasiparticle band at $T_c$
and the change from a more V-like superconducting gap
to a rather U-like pseudogap above $T_c$
can be explained in a consistent way 
by assuming that the low energy pseudogap 
in the underdoped cuprates is due to
phase fluctuations of a local $d_{x^2-y^2}$-wave pairing
gap with fixed magnitude.
Furthermore, phase fluctuations can explain why the 
pseudogap starts closing from the nodal points,
whereas it rather fills in along the anti-nodal directions.

\section*{Acknowledgments}

We would like to acknowledge useful discussions and comments by
E.~Arrigoni, D.~J.~Scalapino and S.~A.~Kivelson. We are grateful to H. Berger for providing us with the high-quality single crystals.
This work was supported by the  KONWHIR projects OOPCV and CUHE
and by the Forschergruppe under Grant No.~FOR 538.
The calculations were carried out at the high-performance computing centers 
HLRS (Stuttgart) and LRZ (M\"unchen).

\bibliographystyle{prsty}
%%% produced via
%%%aux2bib 3dpso5.aux
%\bibliography{/users/eckl/tex/bibtex/references_database,/users/eckl/dissertation/doktorarbeit/references_new.bib}

\end{document}